\documentclass[conference]{IEEEtran}
\IEEEoverridecommandlockouts
\usepackage{graphicx}
\usepackage{booktabs}
\usepackage{amsmath}
\usepackage{amssymb}
\usepackage{url}
\usepackage{adjustbox}
\usepackage{xcolor}
\usepackage{soul}

\def\BibTeX{{\rm B\kern-.05em{\sc i\kern-.025em b}\kern-.08em
		T\kern-.1667em\lower.7ex\hbox{E}\kern-.125emX}}
\begin{document}
\title{When Not to Write Memory: Governing False Promotion from Correlated Agent Traces}

\author{%
\IEEEauthorblockN{Yijiashun Qi$^{*}$}
\IEEEauthorblockA{\textit{University of Michigan}\\
Ann Arbor, MI, USA\\
$^{*}$Corresponding author:\\
elijahqi@umich.edu}
\and
\IEEEauthorblockN{Xiang Xu}
\IEEEauthorblockA{\textit{ByteDance Inc.}\\
San Jose, CA, USA\\
xuxiang.victor@bytedance.com}
\and
\IEEEauthorblockN{Yuxuan Li}
\IEEEauthorblockA{\textit{University of Pennsylvania}\\
Philadelphia, PA, USA\\
yuxuanli@alumni.upenn.edu}
}

\maketitle

\begin{abstract}
Long-lived language agents increasingly write reusable memories from their own execution traces. The key safety question is not only what agents should remember, but when they should refuse to write memory at all. Repeated observations across agents are not necessarily independent evidence: the same claim may be copied from a shared source, induced by a shared prompt, stale under a new environment, or valid only in a narrower scope. We study this failure mode as a memory write-path governance problem. We introduce GovMem as a conservative diagnostic reference policy that estimates dependency-aware support, retrieves counterevidence, assigns scope, and outputs one of three decisions: promote, reject, or needs-review. In controlled synthetic stress tests, GovMem reduces false promotion from 0.597 to 0.040 in the default setting while preserving 0.960 recall, at an explicit review burden. In a project-internal 120-candidate human-labeled real-trace subset spanning 79 recorded traces and project reports, dependency-aware promotion reduces false promotion from 0.371 for source+scope to 0.032 overall, but held-out false promotion remains 0.111 and the method is highly conservative, with 0.692 review burden and 0.448 direct recall. A final human adjudication of 133 high-impact external coding-agent candidates is more severe: none are safe for automatic promotion, and all 11 verification-gate positives are rejected as boilerplate, shared-tool artifacts, file dumps, or non-reusable debugging traces. These results support GovMem primarily as a diagnostic governance design point, not as a generally validated or efficient automatic memory writer: agent memory write paths should be evaluated as risk-controlled evidence-governance systems, while broader external coverage and downstream harm evidence are still needed before stronger claims.
\end{abstract}


\section{Introduction}

Agent memory systems are becoming writable. Modern agents do not only retrieve facts from a static store; they summarize traces, remember user preferences, distill procedures, and reuse lessons from prior tasks. This creates a write-path safety problem: when should an observation in an execution trace become long-term memory, and when should the system refuse to write it?

A central failure mode is that repetition is not evidence. Five agents repeating the same claim may represent five independent discoveries, but they may also represent one copied stale note echoed through shared context. More generally, similar textual observations from multiple agents are tempting to store as reusable experience, yet their confidence is hard to calibrate directly: agreement may come from shared prompts, shared tools, shared intermediate context, or stale scope rather than independent support. If such repeated claims are promoted naively, long-term memory becomes a persistence layer for correlated errors.

This paper studies governed memory promotion from correlated multi-agent traces as a question of when \emph{not} to write memory. We ask: when does repetition become misleading, how should systems govern similar cross-agent observations whose confidence cannot be read off from text alone, and what review burden is required to keep unsafe candidates out of long-term memory?

Our contributions are:
\begin{enumerate}
    \item We identify false promotion from copied, prompt-correlated, stale, adversarial, and out-of-scope observations as a distinct memory failure mode.
    \item We formulate memory writing as an evidence-governance decision rather than a summarization problem, with candidate memories routed to promote, reject, or needs-review.
    \item We instantiate this formulation as GovMem, a conservative side-agent reference policy and diagnostic governance policy that audits write proposals using dependency-aware support, counterevidence, scope, and verification.
    \item We provide controlled mechanism experiments, a project-internal human-gold real-trace evaluation, and external coding-agent stress tests including a human-adjudicated high-impact packet, while explicitly separating completed evidence from local pre-adjudication and scaffolded evaluations and reporting where automatic promotion fails.
\end{enumerate}

\paragraph{Evidence status.}
The controlled GovMem-Bench results are mechanism evidence: they isolate when repetition, source count, and dependency-aware support diverge. The real-trace result is a project-internal evaluation with 120 human-gold candidate memories across 79 recorded traces plus project reports, including copied-source, shared-prompt, shared-tool, low-trust, and unknown dependency cases. The external outcome pilots and the V2 coding-agent stress set test whether the risk-control story survives outside the project scaffold. The strongest new external result is a final human adjudication of a 133-row high-impact V2 packet: it rejects every automatic-promotion candidate, including all 11 local verification-gate positives. This supports a diagnostic safety claim, but not independent external validation of an efficient automatic memory writer. We therefore frame the current manuscript as a diagnostic and method paper, with broader external coverage and downstream-harm evaluation as the main path to stronger claims. Table~\mbox{\ref{tab:evidence-status}} summarizes the evidence status of each evaluation component.

\begin{table*}[t]
\centering
\caption{Evidence status in the current manuscript. Completed, locally pre-adjudicated, human-adjudicated, and scaffolded evaluations are separated so the paper does not overclaim external validity or downstream-harm evidence.}
\begin{adjustbox}{width=0.9\linewidth}
\begin{tabular}{lll}
\toprule
Evidence & Status & Claim supported \\
\midrule
Controlled benchmark & completed & mechanism: repetition can mislead \\
Internal real traces & completed human gold & project-scoped safety tradeoff \\
External outcome pilots & completed semi-auto & pipeline viability; naive baselines over-promote \\
External V2 pool & local pre-adjudication & stress-test construction; local positives not final gold \\
External V2 high-impact packet & completed human adjudication & 0/133 safe automatic promotions; 11/11 gate positives rejected \\
Downstream harm pilot & scaffolded & ready for judgment, no harm result yet \\
\bottomrule
\end{tabular}
\end{adjustbox}
\label{tab:evidence-status}
\end{table*}

\begin{figure}[t]
    \centering
    \includegraphics[width=\linewidth]{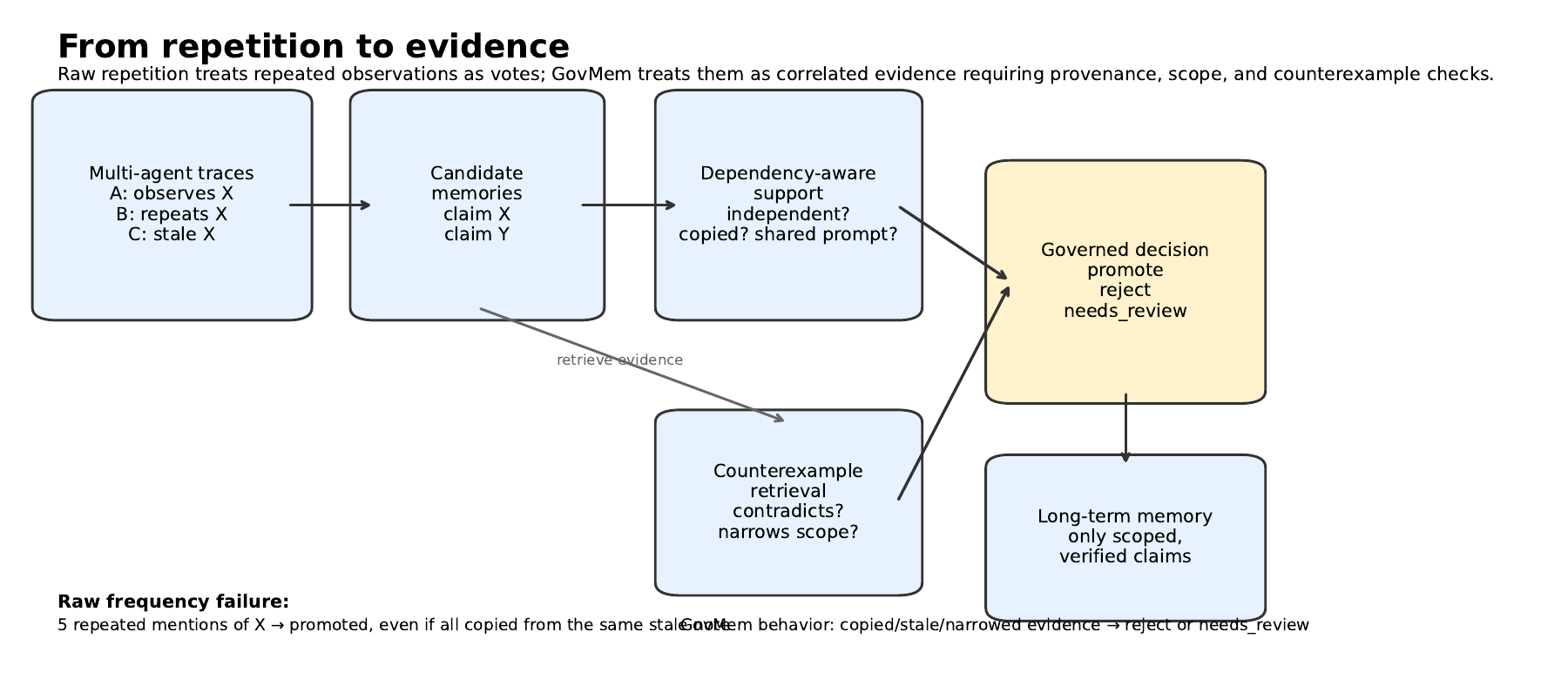}
    \caption{GovMem treats repeated observations as correlated evidence rather than votes. Candidate memories are scored by dependency-aware support, checked against counterevidence, and routed to promote, reject, or needs-review.}
    \label{fig:hero}
\end{figure}

\section{Related Work}

\paragraph{Agent memory systems.}
MemGPT frames LLM memory as an operating-system-like hierarchy for managing context and archival storage~\cite{packer2024memgptllmsoperatingsystems}. A-MEM dynamically organizes long-term agent memories~\cite{xu2025amemagenticmemoryllm}. These systems motivate explicit memory substrates, while GovMem focuses on deciding whether a candidate memory should be written at all.

\paragraph{Reflection and experience distillation.}
Reflexion converts task feedback into verbal reflections~\cite{shinn2023reflexionlanguageagentsverbal}. Generative Agents use observation streams and reflection to simulate believable agents~\cite{park2023generativeagentsinteractivesimulacra}. ExpeL extracts reusable lessons from trajectories~\cite{zhao2024expelllmagentsexperiential}. GovMem differs by treating repeated lessons as evidence requiring provenance, scope, and counterexample checks.

\paragraph{Procedural and multi-agent memory.}
Work such as LEGOMem distills multi-agent trajectories into reusable procedural memories~\cite{han2025legomemmodularproceduralmemory}. GovMem is complementary: it asks when such reusable memories are safe to promote under correlated support.

\paragraph{Governance, provenance, and safety.}
A-MemGuard studies memory poisoning defenses~\cite{wei2025amemguardproactivedefenseframework}. TierMem provides provenance-aware tiered memory with verified fallback to raw evidence~\cite{zhu2026lossyverifiedprovenanceawaretiered}. SSGM proposes a broad governance framework for evolving memory with consolidation and access controls~\cite{lam2026governingevolvingmemoryllm}. GovMem is narrower: it operationalizes one specific write-path failure mode, namely false promotion from correlated multi-agent evidence, and evaluates it with an explicit promote/reject/review benchmark.

\paragraph{Memory benchmarks.}
LoCoMo~\cite{maharana2024evaluatinglongtermconversationalmemory} and MemoryAgentBench~\cite{hu2026evaluatingmemoryllmagents} evaluate long-term memory capabilities such as retention, retrieval, and multi-turn updating. GovMem-Bench instead targets write-time promotion quality under repeated but non-independent observations, where the main failure is not forgetting but wrongly committing unsupported memories.

\paragraph{Graph-based mining and dependency structure.}
More broadly, graph neural networks and hierarchical mining have been used to capture relational structure in high-dimensional imbalanced data~\cite{qi2025graphneuralnetworkdrivenhierarchical}. GovMem uses a different graph: provenance and dependency among agent observations rather than sample-similarity structure.

\paragraph{Alternative governance policies.}
GovMem is not the only plausible way to govern memory writes. A strong LLM judge can directly classify candidate memories from evidence metadata, and a simple review-all-correlated policy can route all non-independent evidence to human review. Our experiments treat these as serious alternatives rather than strawmen: the full-split GPT-5.5 judge has much higher recall and lower review burden than GovMem but higher false promotion, while review-all-correlated nearly matches internal safety with better recall. GovMem should therefore be read as one point in a risk-control design space, emphasizing explicit provenance and conservative write-path decisions rather than universal dominance.

\section{Problem Formulation and GovMem}

Let a trace bundle contain observations $o_i$ from one or more agents. A candidate memory $m$ aggregates observations with similar claim text. Each observation has provenance metadata such as source, prompt family, parent event, environment, and trust. A promotion policy maps $m$ to one of three decisions: $\{\textsc{Promote}, \textsc{Reject}, \textsc{NeedsReview}\}$.

GovMem can be implemented as a conservative side agent: a task-independent write-path auditor that observes raw traces and provenance, estimates whether repeated observations provide independent support, and routes candidate memories to promote, reject, or review. This gives procedural independence from the task agent, but not evidential independence by itself; the side agent must inspect event-level lineage rather than only post-hoc summaries. In our reference design, task agents emit traces, a memory proposer normalizes candidate memories, and GovMem audits each proposed write before it reaches long-term storage.

GovMem makes this decision in four steps. First, it forms a candidate memory by aggregating similar observations while retaining provenance metadata. Second, it estimates effective support by grouping observations according to dependency structure rather than raw count or naive source deduplication: observations that share prompts, tools, intermediate context, or parent events should not count as independent votes. Third, it retrieves and classifies counterevidence while checking target-scope compatibility, since textual agreement alone is not a calibrated confidence signal. Fourth, it applies dependency-aware verification and outputs either promote, reject, or needs-review, making copied or prompt-correlated candidates harder to automatically promote. Figure~\mbox{\ref{fig:hero}} illustrates this write-path pipeline.

\begin{table*}[t]
\centering
\caption{Human annotation unit and fields. Annotators label candidate memories, not every raw trace token. A yes label means the memory is safe to write in the stated scope; no means it would mislead or overclaim; needs-review means the claim may be useful but is not safe for automatic promotion because evidence independence, scope, trust, or counterexamples remain unresolved.}
\begin{adjustbox}{width=0.9\linewidth}
\begin{tabular}{lll}
\toprule
Field & Values or content & Purpose \\
\midrule
Candidate & normalized reusable claim & proposed memory write \\
Evidence refs & trace, event, or document refs & auditability \\
Dependency & independent, copied source, shared prompt/tool, echo, low trust, unknown & estimate effective support \\
Validity & valid, invalid, scope-limited, stale, uncertain & judge truth within scope \\
Scope & task, repo, version, environment, preconditions & prevent overgeneralization \\
Counterexamples & contradicting or narrowing refs & avoid blind promotion \\
Decision & yes, no, needs-review & gold write-path label \\
Rationale & short human explanation & inspectable audit record \\
\bottomrule
\end{tabular}
\end{adjustbox}
\label{tab:annotation-protocol}
\end{table*}

\paragraph{Human labeling protocol.}
Annotators receive the normalized candidate claim, supporting trace or document references, dependency metadata, scope fields, and any retrieved counterexample or narrowing evidence. The operational definition of \emph{safe automatic promotion} is deliberately strict: a candidate can be labeled yes only when the evidence supports the claim, the support is independent or explicitly verified, the reusable scope is clear, no strong counterexample is visible, and writing the memory would not mislead a future agent. Annotators label no when a candidate is contradicted, stale, over-broad, based on failed or low-trust evidence, or merely boilerplate, task-local narration, source dumps, or non-reusable trace content. They label needs-review when the claim may be useful but evidence independence, scope, trust, or external context is unresolved. Disagreements are adjudicated at the row level by inspecting the cited evidence and choosing the narrowest safe non-misleading label; in the final external packet, disputed positive or review labels are not upgraded to yes unless both the reusable claim and its evidence path survive this check. Table~\mbox{\ref{tab:annotation-protocol}} summarizes the per-candidate annotation fields used during labeling.

For the 133-row external high-impact packet, two reviewers independently labeled all rows before adjudication. Exact agreement on the three-way promotion label was 78/133 (58.6\%), with Cohen's $\kappa=0.208$; agreement on dependency type was 56.4\% with $\kappa=0.397$. The low promotion-label agreement is itself informative: most disagreements were not about accepting candidates, but about whether suspicious rows should be rejected outright or routed to review. There were zero two-reviewer consensus yes rows. After resolving 55 disagreements, the final adjudicated labels are 114 no and 19 needs-review, with no safe automatic promotions. Examples of disputed or rejected cases include pytest and warning boilerplate mistaken for verification, task-local debugging narration, file listings or source dumps summarized as reusable knowledge, and patch-success claims that were read-only, failed, or lacked independent outcome evidence.

\section{Controlled GovMem-Bench Experiments}

\paragraph{Benchmark.}
GovMem-Bench generates trace bundles with true independent, true copied, false copied, false correlated, stale, adversarial, and out-of-scope candidate memories. It also supports clustering noise, where extraction merges or splits candidate memories.

\paragraph{Baselines.}
We compare raw frequency, unique-source counting, source+scope gating, confidence-like scoring, counterexample-only promotion, GovMem-lite, and GovMem-verified.

\begin{figure}[t]
    \centering
    \includegraphics[width=0.9\linewidth]{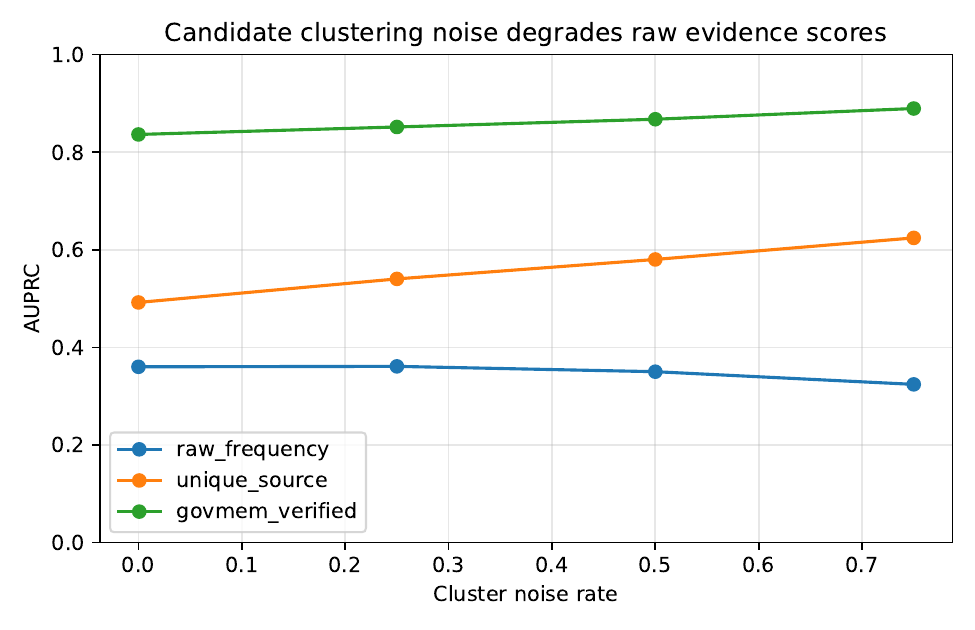}
    \caption{AUPRC under increasing candidate-clustering noise. GovMem remains better calibrated than raw frequency and unique-source counting, indicating that valid memories continue to rank ahead of traps even as clustering quality degrades.}
    \label{fig:cluster-noise}
\end{figure}

\paragraph{Metrics.}
We report precision, recall, false-promotion rate, review burden, budgeted downstream utility, and AUPRC. Precision/recall and false-promotion rate measure write-time safety, review burden measures how much uncertainty is pushed to humans, and B@k/AUPRC measure whether valid memories are surfaced early and ranked ahead of traps. Table~\mbox{\ref{tab:main}} reports the default GovMem-Bench results across all baselines, and Figure~\mbox{\ref{fig:cluster-noise}} shows AUPRC under increasing candidate-clustering noise.

\begin{table*}[t]
\centering
\caption{Default GovMem-Bench results. Raw repetition attains perfect recall but high false promotion. GovMem-verified lowers false promotion from 0.597 to 0.040 while preserving recall at an explicit review cost.}
\begin{adjustbox}{width=0.8\linewidth}
\begin{tabular}{lrrrrrrrr}
\toprule
Method & Precision & Recall & False Prom. & Review & Downstream & B@1 & B@3 & AUPRC \\
\midrule
confidence & 0.271 & 0.372 & 0.729 & 0.000 & 0.200 & 0.613 & 0.200 & 0.491 \\
counterexample & 0.466 & 1.000 & 0.534 & 0.000 & 0.200 & 0.424 & 0.200 & 0.364 \\
govmem-lite & 1.000 & 0.500 & 0.000 & 0.000 & 1.000 & 1.000 & 1.000 & 0.754 \\
govmem-verified & 0.960 & 0.960 & 0.040 & 0.151 & 0.936 & 1.000 & 0.936 & 0.836 \\
raw-frequency & 0.403 & 1.000 & 0.597 & 0.000 & 0.200 & 0.424 & 0.200 & 0.360 \\
source+scope & 0.500 & 0.500 & 0.500 & 0.000 & 0.200 & 0.613 & 0.200 & 0.544 \\
unique-source & 0.375 & 0.500 & 0.625 & 0.000 & 0.200 & 0.613 & 0.200 & 0.492 \\
\bottomrule
\end{tabular}
\end{adjustbox}
\label{tab:main}
\end{table*}

\section{Preliminary Non-Oracle Evaluation}

\begin{figure}[t]
    \centering
    \includegraphics[width=0.48\linewidth]{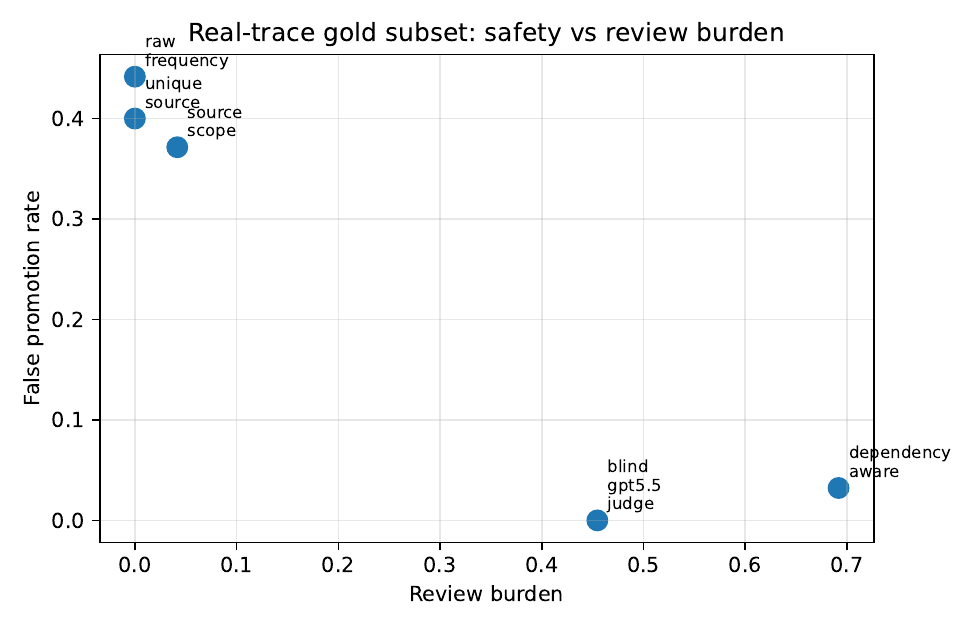}
    \includegraphics[width=0.48\linewidth]{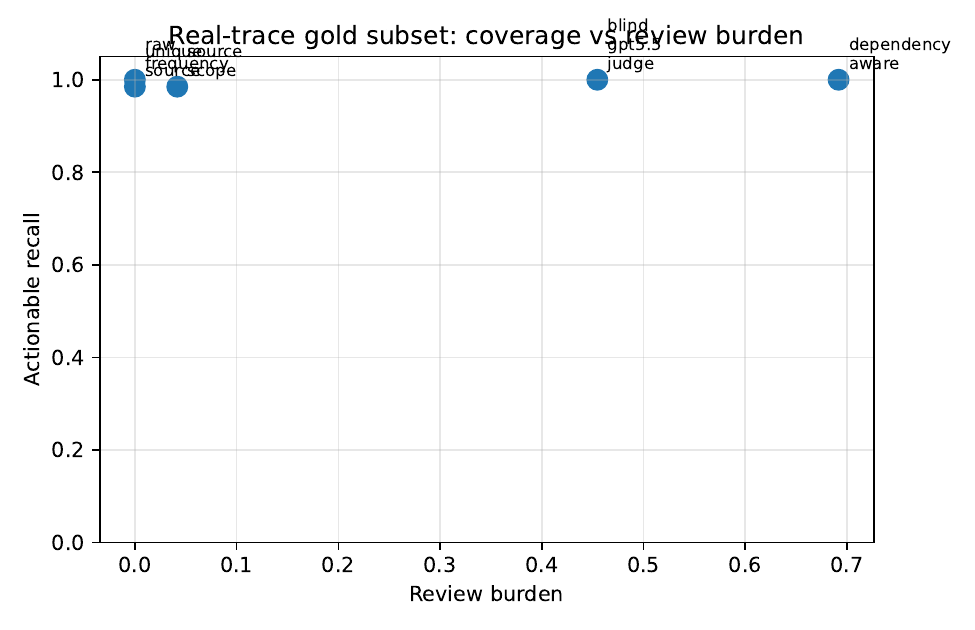}
    \caption{Real-trace gold-subset tradeoffs. Left: false promotion versus review burden. Right: actionable recall versus review burden. On the held-out split, dependency-aware promotion lowers false promotion to 0.111 [0.000, 0.375] versus 0.400 [0.226, 0.581] for source+scope, but only by reviewing many more candidates.}
    \label{fig:real-trace-tradeoff}
\end{figure}

We also build a preliminary real-trace evaluation pipeline: trace schema, candidate extraction from markdown and trace JSON, GPT-assisted pre-labeling, a small human-gold subset, and retrieved counterexample classification. This evaluation contains 120 human-gold candidate memories derived from 79 recorded traces plus project-internal reports, with an 85/35 development/held-out split. It includes copied-source, shared-prompt, shared-tool, low-trust, and unknown dependency cases, but it is still intentionally presented as a project-internal evaluation rather than a powered external comparison. We report direct recall over automatically promoted positive memories and actionable recall over positives that are either promoted or routed to needs-review for scoped human adjudication. Bootstrap intervals over 1000 resamples show the same qualitative tradeoff in both settings: source+scope is less conservative but over-promotes more often, while dependency-aware promotion sharply reduces false promotion by routing many candidates to review. Numerically, the core internal false-promotion comparison is source+scope 0.371 [0.282, 0.467] versus dependency-aware 0.032 [0.000, 0.100]; held-out intervals are wider at 0.400 [0.226, 0.581] versus 0.111 [0.000, 0.375]. Table~\mbox{\ref{tab:real-trace}} reports the full project-internal real-trace numbers, and Figure~\mbox{\ref{fig:real-trace-tradeoff}} visualizes the safety--review tradeoff on the held-out split.

\begin{table*}[t]
\centering
\caption{Project-internal human-gold real-trace subset ($n=120$ candidate memories from 79 recorded traces plus project reports). Dependency-aware promotion is safer than source+scope, but not more efficient: it lowers false promotion from 0.371 to 0.032 while raising review burden from 0.042 to 0.692 and lowering direct recall from 0.985 to 0.448. The full-split GPT-5.5 judge is a strong higher-recall alternative with lower review burden but higher false promotion. This subset is diagnostic and not powered to establish broad superiority.}
\begin{adjustbox}{width=0.9\linewidth}
\begin{tabular}{lrrrrrrr}
\toprule
Method & Promoted & Review & Precision & Direct Recall & Action Recall & False Prom. & Review burden \\
\midrule
raw-frequency & 120 & 0 & 0.558 & 1.000 & 1.000 & 0.442 & 0.000 \\
unique-source & 110 & 0 & 0.600 & 0.985 & 0.985 & 0.400 & 0.000 \\
source+scope & 105 & 5 & 0.629 & 0.985 & 0.985 & 0.371 & 0.042 \\
dependency-aware & 31 & 83 & 0.968 & 0.448 & 1.000 & 0.032 & 0.692 \\
full-split GPT-5.5 judge & 72 & 45 & 0.903 & 0.970 & 1.000 & 0.097 & 0.375 \\
\bottomrule
\end{tabular}
\end{adjustbox}
\label{tab:real-trace}
\end{table*}

A strong review-budget baseline, \emph{review-all-correlated}, performs surprisingly well: it routes every non-independent dependency type to review and applies source+scope only to independent candidates. On all labels it nearly matches dependency-aware false promotion (0.033 versus 0.032), while achieving higher recall (0.881 versus 0.448) and lower review burden (0.492 versus 0.692). This makes the empirical story deliberately modest. GovMem's current advantage is over raw, source-count, and random-review policies; it does not yet dominate simple correlated-evidence review. The right reading is that GovMem is one design point in a risk-control space. It may be preferable when systems need structured audit records, explicit counterexample retrieval, scope assignment, and dependency categories that can later support selective automation. The simpler baseline may be preferable when the deployment can afford to review every non-independent candidate and only needs a conservative triage rule.

This category-level view clarifies that GovMem's safety comes primarily from routing dependency-risky candidates to review rather than from automatically promoting more memories; therefore, high actionable recall can coexist with low direct recall. Table~\mbox{\ref{tab:dependency-breakdown}} reports false promotion by dependency type on the 83-candidate diagnostic subset, Table~\mbox{\ref{tab:candidate-audit}} lists representative candidate-level audit rows, and Table~\mbox{\ref{tab:labeling-ladder}} situates this evaluation on a practical human-labeling ladder.

\begin{table*}[t]
\centering
\caption{Practical human-labeling ladder for this problem. The current internal evaluation, with 120 human-gold candidates across 79 traces, reaches the workshop/weak-main diagnostic scale. Stronger claims require not just more internal labels, but independent adjudication of external candidates and executed downstream-harm judgments.}
\begin{adjustbox}{width=0.9\linewidth}
\begin{tabular}{llll}
\toprule
Evidence level & Human labels & Traces & Claim supported \\
\midrule
Workshop diagnostic & 80--120 & 30--50 & failure mode and benchmark viability \\
Weak-main diagnostic & 120--200, with double-coding & 50--100 & conservative safety-filter tradeoff \\
Main-conference evidence & 300--500, including 100+ external & 100+ & independent external validity plus harm evidence \\
\bottomrule
\end{tabular}
\end{adjustbox}
\label{tab:labeling-ladder}
\end{table*}

\begin{table*}[t]
\centering
\caption{Practical human-labeling ladder for this problem. The current internal evaluation, with 120 human-gold candidates across 79 traces, reaches the workshop/weak-main diagnostic scale. Stronger claims require not just more internal labels, but independent adjudication of external candidates and executed downstream-harm judgments.}
\begin{adjustbox}{width=0.75\linewidth}
\begin{tabular}{lrrrrrrr}
\toprule
Dependency & $n$ & Raw FP & S+S FP & DA Prom. & DA Rev. & DA FP & DA Action Rec. \\
\midrule
copied-source & 3 & 1.000 & 0.000 & 0 & 3 & 0.000 & 0.000 \\
independent & 46 & 0.152 & 0.049 & 16 & 25 & 0.000 & 1.000 \\
low-trust & 8 & 1.000 & 1.000 & 0 & 8 & 0.000 & 0.000 \\
same-agent echo & 7 & 0.714 & 0.667 & 2 & 4 & 0.000 & 1.000 \\
shared-prompt & 4 & 1.000 & 1.000 & 0 & 4 & 0.000 & 0.000 \\
shared-tool & 8 & 0.500 & 0.429 & 3 & 5 & 0.000 & 1.000 \\
unknown & 7 & 1.000 & 1.000 & 0 & 7 & 0.000 & 0.000 \\
\bottomrule
\end{tabular}
\end{adjustbox}
\label{tab:dependency-breakdown}
\end{table*}

\begin{table*}[t]
\centering
\caption{Representative candidate-level audit rows. GovMem avoids several unsafe promotions that simpler baselines would accept, but it also routes some valid project-scoped positives to review rather than automatically promote them. The table is diagnostic: it makes the safety--review tradeoff concrete at the candidate level instead of only through aggregate metrics.}
\begin{adjustbox}{width=0.9\linewidth}
\begin{tabular}{lllll}
\toprule
Candidate type & Example claim & Gold & Source+scope & GovMem \\
\midrule
Hard negative & Synthetic results are sufficient top-tier evidence & reject & review & reject \\
Low-trust evidence & Generated trace warnings support real-world performance & review & promote & review \\
Shared prompt & Same-prompt reviewers identify current blocker & review & promote & review \\
Reviewed positive & Unique-source remains a strong baseline & promote & promote & review \\
Reviewed positive & LLM extraction fidelity is a bottleneck & promote & promote & review \\
\bottomrule
\end{tabular}
\end{adjustbox}
\label{tab:candidate-audit}
\end{table*}

\begin{table*}[t]
	\centering
	\caption{Final human adjudication of the 133-row high-impact V2 external packet. The result overturns the local positive labels: zero candidates are safe for automatic promotion, and all 11 verification-gate positives are rejected. This strengthens the diagnostic claim that surface verification and repeated trace evidence are not enough for safe memory writes.}
	\begin{adjustbox}{width=0.75\linewidth}
		\begin{tabular}{lrrrrl}
			\toprule
			High-impact slice & $n$ & Yes & Review & No & Main failure mode \\
			\midrule
			Verification-gate positives & 11 & 0 & 0 & 11 & shared-tool boilerplate \\
			Dependency-aware promotions & 72 & 0 & 17 & 55 & failed/read-only patch claims \\
			Borderline candidates & 50 & 0 & 2 & 48 & file dumps or task-local narration \\
			\midrule
			Total & 133 & 0 & 19 & 114 & no safe automatic promotion \\
			\bottomrule
		\end{tabular}
	\end{adjustbox}
	\label{tab:external-human-adj}
\end{table*}

\section{External Coding-Agent Stress Tests and Failure Analysis}

As secondary external stress tests, we convert public SWE-agent and OpenHands trajectories into the GovMem trace schema. In two outcome-grounded pilots, each with 50 trajectories and 100 candidate labels, raw frequency, unique-source, and source+scope each have false promotion 0.500, while dependency-aware promotion has precision 1.000, recall 0.500, actionable recall 1.000, and false promotion 0.000 at review burden 0.750. Combined across both scaffolds, this yields 100 external trajectories and 200 outcome-grounded labels with the same qualitative pattern. We use these pilots to show that the external pipeline runs end-to-end and that naive promotion policies can over-promote across scaffolds; we do not treat them as broad external validation.

We also construct a harder content-grounded external stress set from the same SWE-agent and OpenHands samples. After filtering boilerplate-like excerpts, the V2 pool contains 361 candidates. Before final human review, local conservative adjudication had labeled 11 candidates as promote-worthy, 300 as needs-review, and 50 as reject; the earlier dependency-aware policy promoted 72 candidates that were already suspicious under the local stress analysis. This local curve was useful for selecting high-impact rows, but it was not final gold.

Final human adjudication is the central external result, not a secondary check. We double-review and resolve a 133-row high-impact packet consisting of the 11 local verification-gate positives, 72 dependency-aware false-promotion candidates, and 50 borderline candidates. The final labels contain no safe automatic promotions: 114 candidates are rejected and 19 require review. Most importantly, all 11 verification-gate positives are rejected. Their apparent verification evidence is usually pytest boilerplate, warnings, file listings, source dumps, failed edits, or the agent's own debugging narration rather than a reusable verified memory. Thus the apparent low-coverage safe operating point in the local V2 curve was an adjudication artifact, not an external success case.

The failures cluster into four recurring categories. \emph{Boilerplate verification} cases confuse standard pytest output, warning banners, or command echoes with evidence that the candidate memory itself was verified. \emph{Task-local narration} cases summarize what the agent attempted in one issue, but the statement would not transfer safely to future tasks. \emph{File dumps} convert source listings or repository inventory into broad procedural memories without demonstrating correctness. \emph{Non-reusable debugging traces} preserve failed edits, read-only inspections, or speculative diagnoses as if they were durable project knowledge. These categories explain why local automatic positives collapse under human review and motivate stronger event-level evidence extraction before any external automatic-promotion claim. Table~\mbox{\ref{tab:external-human-adj}} summarizes the final human adjudication of the 133-row high-impact V2 packet by failure mode and decision.

This negative result is central to the revised external claim. GovMem should not be described as an externally validated automatic promotion method; instead, the external evidence supports a narrower diagnostic thesis that memory write paths need stricter evidence extraction, provenance checks, and review routing. The current external safe-coverage result is effectively zero on the human-adjudicated high-impact packet, and recovering useful automatic coverage without false promotion remains future work.

Finally, we prepare a downstream memory-harm pilot to test whether false promoted memories alter later agent judgments. The pilot contains 35 prompt-level judgment tasks: 12 false-memory injected cases, 12 review-gate controls, and 11 safe-memory positive controls. The prompts and scoring templates are ready, but no LLM or human judgments have been run, so this remains an experiment scaffold rather than harm evidence.

\section{Limitations and Threats to Validity}

The main controlled benchmark is synthetic and aligned with the method's intended signals. The real-trace subset is still project-internal: it contains 120 human-gold candidate memories from 79 recorded traces and project reports, with 35 held out. The external SWE-agent and OpenHands outcome pilots are semi-automatically labeled, and although the V2 high-impact packet now has final human adjudication, it is a selected 133-row stress slice rather than a full independently labeled external benchmark. The remaining 361-candidate V2 pool still contains local pre-adjudication labels. The downstream harm experiment is scaffolded but not executed. Counterexample retrieval is lightweight and report-based rather than a strong retrieval benchmark.

Several claim boundaries follow from these limitations. First, GovMem should not be read as an efficient automatic memory writer: the internal dependency-aware policy trades safety for high review burden, and the external verification-selective gate fails under final human review because surface verification cues can be boilerplate or non-reusable trace artifacts. Second, GovMem does not dominate simple review-all-correlated baselines in the current internal evaluation, so the contribution is the diagnostic governance framework and risk-controlled write-path framing, not blanket empirical superiority. Third, the external V2 result is best read as a stress test of failure under harder candidate quality rather than as a positive deployment result: both dependency-aware promotions and verification-gate positives collapse to zero human-approved automatic promotions in the high-impact packet. Taken together, these limitations place the current paper in the weak-main diagnostic range rather than the full main-conference range.

\section{Conclusion}

Repeated agent observations are not independent votes, and many repeated claims should not be written to memory. GovMem reframes memory writing as evidence governance: candidate memories require dependency-aware support, counterevidence, scope, and an explicit review path. The current evidence supports a diagnostic safety-filter claim: governed write paths can reduce false promotion internally, but external human adjudication shows that automatic promotion remains brittle when verification evidence is extracted from noisy agent traces. Stronger claims will require broader independently labeled external coverage, executed downstream-harm tests, and stronger event-level verification.

\bibliographystyle{ieeetr}
\bibliography{references}

\end{document}